\documentstyle[12pt]{article}
\newcommand{\bn}[1]{\mbox{\boldmath$#1$}}

\newcommand{\beq}{\begin{equation}}
\newcommand{\eeq}{\end {equation}}
\newcommand{\bea}{\begin{eqnarray}}
\newcommand{\eea}{\end{eqnarray}}

\begin{document}
\lineskip=24pt
\baselineskip=24pt
\raggedbottom

{\Large{\centerline{\bf Quantum theory of spontaneous emission}}}
{\Large{\centerline{\bf by real moving atoms}}}
\vskip 1.5cm
\centerline{{\bf L. G. Boussiakou, C. R. Bennett} and {\bf M.
Babiker}}
\centerline{ Department of Physics, University of York, Heslington,
York YO10
5DD, England}
\vskip 2.0cm
\section*{Abstract} We outline the solution of a fundamental problem in quantum
theory which has hitherto lacked a proper solution, namely finding  the requisite quantum theoretical framework 
guaranteeing that the calculated inverse spontaneous emission rate of a moving atom, as a composite system of charged particles
interacting with the Maxwell field, is slowed  down exactly as in time dilation.

\vskip 2.5cm
\centerline{\bf PACS numbers: 03.30.+p, 31.30.Jv, 42.50.Ct, 03.65.Vf}
\newpage
It has recently been shown that a neutral atom bearing an  
electric dipole moment moving in an external magnetic field can accumulate 
a quantum phase [1-5].  A moving dipole may,  under suitable conditions, exhibit a detectable Aharonov-Bohm phase shift [6] 
and the rotational motion of a Bose-Einstein condensate in a vortex state can induce a  magnetic monopole [7,8] distribution
 or an electric charge distribution [8]. These and other effects  associated with atomic motion continue to receive considerable attention and especially so with the
advent of atom optics [9,10] and laser cooling and trapping  [11].
At first sight it would appear that the requisite theory for the description of pheneomena involving moving atoms could be 
constructed as a straightforward extension of non-relativistic quantum optics by incorporating the translational motion of 
the atomic centre of mass.

In fact the need to incorporate the centre of mass motion in quantum optics theory 
had necessitated a re-appraisal of the corresponding non-relativistic quantum electrodynamical theory 
where investigations  sought to elucidate how the division of the motion into centre of mass and 
internal motions is affected by the presence of the interaction with electromagnetic fields [12-15].  One of
the main outcomes of these investigations was the emphasis on the role of the 
R\"{o}ntgen interaction [16] energy term which couples the electric dipole moment to an
effective electric field involving the centre of mass velocity and the magnetic
field of the light.  

It was Wilkens [17] who pointed out that a theory which excluded the R\"{o}ntgen interaction 
would lead to spurious velocity-dependent effects when evaluating the spontaneous decay rate of 
an excited electric dipole moving freely in electromagnetic vacuum.  Wilkens extended
his work to include the R\"{o}ntgen interaction and evaluated the
scattering rate into a given solid angle in a given direction,
deducing that this was free of any spurious velocity-dependences [18].
He did not,  however,
proceed to ascertain whether the calculated {\underline {total}}
spontaneous emission rate based on his approach would be consistent with the 
requirements of special relativity.  It turns out that Wilken's approach  
in fact leads to an incorrect result for the total rate. 

More recently, Barton  and Calegoracos [19] highlighted the absence in the
literature of a proper treatment 
of the quantum theory of spontaneous emission of atoms moving in a
classically assigned
trajectory.  This is so even for the simplest case of a uniformly moving
atom.  They put forward a theory in which they considered a model system of an atom in which the nucleus is 
the centre of mass and merely provides the Coulomb potential
binding it to the electron and this system was assumed to be 
interacting only with the scalar field 
as a simplified representative of the Maxwell field. In addition to these
simplifications, their theory called 
for a careful distinction between energies and Hamiltonians.
As far as the authors are aware, a workable theoretical framework
of the problem in which a real atom interacts with the full (vector) Maxwell
field is hitherto  
unknown and it is our purpose here to furnish such a framework.

Our theory makes use of unambiguous canonical techniques;
Lorentz transformations and a gauge 
transformation are two of its ingredients.  It turns out that not only 
the R\"{o}ntgen interaction features 
prominently in the theory, but that modified atomic internal energy levels
and eigenfunctions, which produce an expected Doppler shift,
play important roles in the characteristics of the emission process.
Finally, one needs to distinguish carefully
between projections of vector fields parallel and transverse to the atomic
velocity vector.
These different aspects of the problem conspire in a remarkable manner
leading to the correct result, namely
that the inverse total spontaneous rate of the moving atom follows the time
dilation formula, as required by special relativity.

The model of a real atom we consider here involves two oppositely charged particles of
charges $e_1=-e_2=e$ and finite masses $m_1$ and $m_2$.  In the centre of mass frame (atomic frame) we denote the position
vectors of the two particles by ${\bf q}'_1$
and ${\bf q}'_2$, and the electromagnetic
scalar and vector potentials as $\phi'({\bf r}')$ and ${\bf A}'({\bf r}')$, respectively.  The Lagrangian density
for the electromagnetic field including the interaction with the two charged particles is
\beq
{\cal L}'=\frac{\epsilon_0}{2}\left[({\bf {\dot A}}'({\bf r}')+
{\bn \nabla}'\phi'({\bf r}'))^2-
c^2({\bn \nabla}'\times {\bf A}'({\bf r}'))^2\right]+
{\bf J}'({\bf r}').{\bf A}'({\bf r}')-\rho'({\bf r}') \phi' ({\bf r}')
\label{2}
\eeq
where the electric field is ${\bf E}'({\bf r}')=-({\bf {\dot A}}'({\bf r}')+
{\bn \nabla}'\phi'({\bf r}'))$, the magnetic field is ${\bf B}'({\bf r}')=
{\bn \nabla}'\times {\bf A'}({\bf r})$ and the current and charge densities
of the particles are given, respectively, by
\beq
{\bf J}'({\bf r}')=e\left[{\bf {\dot q}}'_1 \delta({\bf r}'-{\bf q}'_1)-
{\bf {\dot q}}'_2 \delta({\bf r}'-{\bf q}'_2)\right]
\label{3}
\eeq
\beq
\rho'({\bf r}')=e\left[\delta({\bf r}'-{\bf q}'_1)-
\delta({\bf r}'-{\bf q}'_2)\right]
\label{4}
\eeq
The notation is such  that the atomic frame (the rest-frame) is referred to as $S'$ and
all quantities relative to this frame are primed.  The laboratory frame is the unprimed frame and will be referred
to as $S$,  relative to which the atomic centre of mass moves at velocity
${\bf {\dot R}}$, and all quantities relative to $S$ are unprimed.

The Langrangian density in Eq.(\ref{2}) can be recast in terms of the primed centre of mass coordinates, defined by
\beq
{\bf R}'=\frac{(m_1 {\bf q}'_1 + m_2 {\bf q}'_2)}{M};\;\;\;\;\;\;\;
{\bf q}'={\bf q}'_1 - {\bf q}'_2
\label{5}
\eeq
where $M=m_1+m_2$ is the atomic mass. Note that in the primed frame,
or rest-frame, we must have ${\bf {\dot R}}'=0$,
{\it by definition}.  This allows us to carry out a Power-Zienau-Woolley
gauge transformation [20] and straightforwardly obtain the new Langragian
density 
\bea
{\cal L}'=\frac{\epsilon_0}{2}\left[({\bf {\dot A}}'({\bf r}')+
{\bn \nabla}'\phi'({\bf r}'))^2-
c^2({\bn \nabla}'\times {\bf A}'({\bf r}'))^2\right]
-{\bn {\cal P}}'({\bf r}').({\bf {\dot A}'}({\bf r}')+
{\bn \nabla}'\phi'({\bf r}'))
\nonumber \\
+{\bn {\cal M}}'({\bf r}').({\bn \nabla}'\times {\bf A}'({\bf r}'))
\label{6}
\eea
where the polarisation and magnetisation vectors are expressed as full multipolar series
in closed analytical forms 
\beq
{\bn {\cal P}}'({\bf r}')=\sum_{i=1,2} e_i \int^1_0 d\lambda ({\bf q'_i-R'}) \;
\delta({\bf r'-R'}-\lambda({\bf q'_i-R'}))
\label{7}
\eeq
\beq
{\bn {\cal M}}'({\bf r}')=\sum_{i=1,2} e_i \int^1_0 d\lambda
\lambda ({\bf q'_i-R'})\times
{\bf {\dot q}'_i} \; \delta({\bf r'-R'}-\lambda({\bf q'_i-R'}))
\label{8}
\eeq
The Lagrangian density in Eq.(\ref{6}) has a manifestly covariant form, viz
\beq
{\cal L}'=-\frac{\epsilon_0}{4} F'^{\mu\nu}F'_{\mu\nu}-
\frac{1}{2}G'^{\mu\nu}F'_{\mu\nu}
\label{9}
\eeq
where in the primed $S'$ frame $F'_{\mu\nu}$ is the well known
electromagnetic field 4-tensor [21] and
$G'_{\mu\nu}$ is the polarisation field 4-tensor [22].
Formally $G'^{\mu\nu}$ has the same
form as $F'^{\mu\nu}$ but with the substitutions ${\bf E'}\rightarrow
{\bn {\cal P}}'$ and $c{\bf B}'\rightarrow-{\bn {\cal M}}'/c$. 

Lorentz invariance allows us to write the Lagrangian density in the unprimed (laboratory) frame $S$ exactly as in
Eq.(\ref{9}), or Eq.(\ref{6}),  simply by removing the primes.
The total Lagrangian in the unprimed frame
can now be written by adding the familiar relativistic Lagrangian
contributions from the two particles as follows
\beq
L=-m_1{\dot q}_1^2 / \gamma({\dot q}_1)-
m_2{\dot q}_2^2 /\gamma({\dot q}_2)
+\int d^{3}{\bf r} \left[\frac{\epsilon_0}{2}\left(E^2({\bf r})-c^2B^2({\bf r})\right)
+{\bn {\cal P}}({\bf r}).{\bf E}({\bf r})+
{\bn {\cal M}}({\bf r}).{\bf B}({\bf r})\right]
\label{10}
\eeq
where $\gamma({\dot q})=(1-{\dot q}^2/c^2)^{-1/2}$;  the electric and magnetic fields are given by
${\bf E}({\bf r})=-{\bf {\dot A}}({\bf r})-{\bn {\nabla}}\phi({\bf r})$ and ${\bf B}({\bf r})={\bn {\nabla}}\times {\bf A}({\bf r})$.
It is important to bear in mind that the unprimed polarisation and magnetisation fields ${\bn {\cal P}}({\bf r})$ and ${\bn {\cal M}}({\bf r})$ appearing 
in Eq.(\ref{10}) are {\it not} those in Eqs.(\ref{7}) and
(\ref{8}). The primed polarisation and magnetisation fields are rest properties and the unprimed ones are related to 
them by relativistic connection 
rules involving a Lorentz transformation of the polarisation field 4-tensor $G_{\mu\nu}$ [22]. The interaction Lagrangian density
(the last two terms in Eq.(\ref{10})) can thus be rewritten in terms of the primed polarisation
and magnetisation by direct substitution as follows,
\bea
{\cal L}_{int}=\left[{\bn {\cal P}}'_\parallel({\bf r}')+\gamma
\left({\bn {\cal P}}'_\perp({\bf r}')+
\frac{1}{c^2}{\bf {\dot R}}\times
{\bn {\cal M}}'({\bf r}')\right)\right].{\bf E}({\bf r})+
\nonumber \\
\left[{\bn {\cal M}}'_\parallel({\bf r}')+\gamma
\left({\bn {\cal M}}'_\perp({\bf r}')-
{\bf {\dot R}}\times
{\bn {\cal P}}'({\bf r}')\right)\right].{\bf B}({\bf r})
\label{11}
\eea
where the subscript $\parallel$ ($\perp$) denotes the vector projection parallel (perpendicular) to ${\bf {\dot R}}$
and $\gamma=\gamma({\dot R})$.
Note that in Eq.(\ref{11})
the electric and magnetic fields have been left untransformed and we have emphasised the dependence on the space
arguments ${\bf r}$ and ${\bf r'}$ which 
are connected by the Lorentz coordinate transformation equations.
The term in Eq.(\ref{11}) involving the velocity, electric polorisation and the magnetic ${\bf B}$ field is
identified as the R\"ontgen Lagrangian interaction term [12-16], while the term involving the velocity,
magnetisation and the electric field is identified as the Aharonov-Casher
term [8,23].

Having expressed the interaction Lagrangian in terms
of the known rest atomic properties ${\bn {\cal P}}'$ and ${\bn {\cal M}}'$,
we now turn to the particle Lagrangian terms (given by the first two terms in Eq.(\ref{10})) and seek to express their sum in 
terms of the unprimed
centre of mass and relative velocities, ${\bf {\dot R}}$ and
${\bf {\dot q}}$, respectively,  using the relations
\beq
{\bf {\dot q}}_1={\bf {\dot R}}+\frac{m_2}{M}{\bf {\dot q}}; \;\;\;\;\;\;\;
{\bf {\dot q}}_2={\bf {\dot R}}-\frac{m_1}{M}{\bf {\dot q}}
\label{12}
\eeq
Concentrating on the unprimed (S) frame we now  make use of the fact that the internal dynamics 
of the atom are not affected by relativistic considerations other than through ${\bf {\dot R}}$ (i.e. the motion of the electron
round the nucleus is  essentially non-relativistic).
We may then expand the sum of the particle Lagrangian terms up to terms quadratic in ${\bf {\dot q}}$ to obtain
\bea
-\left(m_1{\dot q}_1^2 / \gamma({\dot q}_1)+
m_2{\dot q}_2^2 / \gamma({\dot q}_2)\right)&\simeq&
-M {\dot R}^2 / \gamma + \frac{1}{2}\gamma\mu\left({\dot q}^2
+\frac{\gamma^2}{c^2}({\bf {\dot R}}.{\bf {\dot q}})^2\right)
\nonumber \\
&\simeq&-Mc^2+\frac{1}{2}\left[M{\dot R}^2+\gamma\mu\left({\dot q}^2_\perp
+\gamma^2{\dot q}^2_\parallel\right)\right]
\label{13}
\eea
where $\mu=m_1m_2/M$ is the reduced mass.  

After making use of Eqs.(\ref{11}) and (\ref{13}), the new Lagrangian emerging from Eq.(\ref{10}) 
now becomes the starting point of the canonical procedure with ${\bf R}$ and ${\bf q}$ as the canonical variables for the atom 
and ${\bf A}({\bf r})$ and $\phi({\bf r})$
for the fields. The canonical momenta are
${\bf P}$ (conjugate to ${\bf R}$), ${\bf p}$ (conjugate to ${\bf q}$) and
${\bf \Pi}({\bf r})$, which is identified as $-{\bf D}({\bf r})$,
the electric displacement field, is the momentum conjugate
to ${\bf A}({\bf r})$,  while the momentum conjugate to $\phi$ is zero.
 Since we are interested in spontaneous emission by an atom characterised by an electric dipole moment,  we may now ignore
magnetic interactions by setting all terms involving ${\bn {\cal M}}$ to zero.  The final Hamiltonian emerging from the
canonical procedure can be written as a sum of three terms as follows
\beq
H=H^{0}_{a}+H^{0}_{f}+H_{int}
\eeq
where
\beq
H^{0}_{a}=\frac{P^2}{2M}+\frac{1}{2\gamma\mu}\left(p_\perp^2+
\frac{p_\parallel^2}{\gamma^2}\right)+U(q)\label{14}
\eeq
\beq
H^{0}_{f}=\int d^{3}{\bf r} \frac{\epsilon_0}{2}\left[
\frac{1}{\epsilon_0^2}\Pi^2({\bf r})+c^2B^2({\bf r})\right]\label{15}
\eeq
\bea
H_{int}&=&\int d^{3}{\bf r} \left[\frac{1}{\epsilon_0}
\left({\bn {\cal P}}'_\parallel({\bf r}')+
\gamma{\bn {\cal P}}'_\perp({\bf r}')\right).{\bf \Pi}({\bf r})+\right.
\nonumber\\
&& \left.\frac{{\bf P}}{2M}.({\bn {\cal P}}'({\bf r}')\times{\bf B}({\bf r}))
+({\bn {\cal P}}'({\bf r}')\times{\bf B}({\bf r})).\frac{{\bf P}}{2M}
\right]
\label{16}
\eea
In Eq.(\ref{14}) $H^{0}_{a}$ is identified as the unperturbed atomic Hamiltonian and is seen to be clearly divisible into a centre of mass
part and an internal part and we should note the appearance of the relativistic factor $\gamma$ in the latter.
The potential $U(q)$ in Eq.(\ref{14}) is the inter-particle Coulomb potential in the unprimed (laboratory) frame.  
In the primed frame (rest-frame)
the inter-particle Coulomb potential, denoted as $U'(q')$, arises in the multipolar formulation from an integral term containing the
square of the irrotational part ${\bn {\cal P}}'^{L}$ of the polarisation field, together with infinite Coulomb self energies
 \beq
 \frac{1}{2\epsilon_{0}}\int d^{3}{\bf r'}\left\{{\bn {\cal P}'}^{L}\right\}^{2}=U'(q')+\;{\rm {infinite\; Coulomb\; self\; energies}}
 \eeq
 On disregarding the infinite Coulomb self energies, one then transforms the inter-particle Coulomb energy $U'(q')$ to obtain $U(q)$,  the interaction
 in the unprimed frame. 
The simplest and most direct route is by following the force transformation argument [24] to obtain
\beq
U(q)=\frac{U'(q')}{\gamma}=-\frac{e^2}{4\pi\epsilon_0\gamma q'}
\label{17}
\eeq

The expression for  $H^{0}_{f}$ given in Eq.(\ref{15}) is the familiar unperturbed field Hamiltonian which can be quantised following the standard 
methods of quantisation for a free field in the laboratory frame.  Finally, $H_{int}$,  given in Eq.(\ref{16})
is a new form of interaction Hamiltonian which couples the rest polarisation field to electromagnetic fields in the laboratory frame.
Note the division of polarisation in terms of $\parallel$ and $\bot$ contributions and the appearance of the relativistic 
factor $\gamma$, explicitly and also implicitly via Lorentz transformation from ${\bf r'}$ to ${\bf r}$.  The last set of terms
in $H_{int}$ are identified as the R\"{o}ntgen interaction,  expressed in a symmetrised form. 

In order to describe the spontaneous emission process,  we need to solve zero order
eigenproblems for the fields and for the atom. The fields  can be quantised by following the standard 
methods of quantisation based on the Hamiltonian in Eq.(\ref{15}) in the laboratory frame. 
The solution of the atomic eigneproblem involves consideration of the Schr{\"o}ndinger equation 
$H^{0}_{a}\Psi({\bf R},{\bf q})=E\Psi({\bf R},{\bf q})$ which can be written as
\beq
\left[-\frac{\hbar^2}{2M}\nabla^2_{\bf R}
-\frac{\hbar^2}{2\gamma\mu}\left(\frac{\partial^2}{\partial q_x^2}+
\frac{\partial^2}{\partial q_y^2}+
\frac{\partial^2}{\gamma^2 \partial q_z^2}\right)
-\frac{e^2}{4\pi\epsilon_0\gamma q'} \right] \Psi({\bf R},{\bf q})=
E\Psi({\bf R},{\bf q})
\label{18}
\eeq
where, without loss of generality, we have taken the direction of the velocity ${\bf {\dot R}}$ to be the
$z$-direction. Equation (\ref{18}) admits solutions of the form
$\Psi({\bf R},{\bf q})=e^{i{\bf K}.{\bf R}}\psi({\bf q})$
where $E=\hbar^2 K^2 / 2M+\epsilon$. Upon making the substitution
$q'_z=\gamma q_z$ we obtain a
Sch{\"o}dinger equation governing the internal states of a hydrogenic atom in the atomic frame (rest-frame) $S'$
such that $\gamma \epsilon = \epsilon'$ where $\epsilon'$ are
the internal eigenenergies in the rest-frame $S'$. 

The corresponding eigenfunction possesses the same formal expression in
the two frames, but note that in the laboratory frame $S$ the position vector
would be ${\bf q}$ not ${\bf q}'$ which means that the atom will appear to 
Lorentz-contract in the direction parallel to the velocity, as should be expected.
It is important to remember that these physically consistent modifications to
the internal energy levels and eigenfunctions have only come to light
because of the modified form
of the internal kinetic energy term as well as the Coulomb potential energy
terms appearing in Eqs.(\ref{14}) and (\ref{17}). Without
the asymmetry due to the presence of $\gamma$ in the internal kinetic energy terms and the dependence of the Coulomb potential 
energy on $q'$, the familiar spatial symmetry of the hydrogenic Sch{\"o}ndinger equation
would have been lost,  leading to angular dependence and, consequently, to spurious features arising from the lifting of the
degeneracy of the energy levels. 

We are now
in a position to consider the energy and momentum conservation accompanying 
the process of  
spontaneous emission of a photon described in the unprimed (laboratory) frame
 as having wavevector ${\bf k}$ and frequency $\omega$ when
the internal energy of the atom changes from $\epsilon_i$ to $\epsilon_f$.
Conservation of momentum requires that we have ${\bf K}_f={\bf K}_i-{\bf k}$ where
${\bf K}_i=M{\bf {\dot R}} / \hbar$ is the initial centre of mass wavevector
 and ${\bf K}_f$ is the final wavevector in the laboratory frame.  Conservation of energy, on the other hand,
 demands that we have
\beq
\omega=\frac{1}{\hbar}(\epsilon_i-\epsilon_f)+
\frac{\hbar}{2M}({\bf K}_i-{\bf K}_f)^2
=\frac{1}{\hbar\gamma}(\epsilon'_i-\epsilon'_f)+
\frac{\hbar^2}{M}({\bf K}_i.{\bf k})-\frac{\hbar k^2}{2M}
\simeq \frac{\omega'_0}{\gamma}+{\bf {\dot R}}.{\bf k}
\label{19}
\eeq
where $\hbar\omega'_0$ is the energy level difference in the primed (rest) frame
$S'$ and we have ignored the second order recoil energy. Note that
Eq.(\ref {19}) is equivalent to a Doppler shift in the photon frequency.

Two cases in the calculation of the spontaneous emission rate will have to be considered relative to
the laboratory (unprimed) frame S, namely (i) when
the dipole moment vector is parallel to the velocity vector and (ii) when the dipole meoment vector
 is perpendicular to the velocity vector. If these two calculations yield exactly the same result, then the spontaneous
emission is deemed to be isotropic i.e. free from angular dependence. Imposing the electric dipole approximation,
${\bn {\cal P}}'({\bf r}')={\bf d}'\delta({\bf r}'-{\bf R}')$ where
${\bf d}'=e{\bf q}'$, we obtain for the transition matrix element squared,
with $H_{int}$ as given in Eq.(\ref{16})
as the interaction,
\beq
|ME|^2=d'^2\left|\frac{{\bf E}_\parallel({\bf R})}{\gamma}+
\left[{\bf E}_\perp({\bf R})-\frac{1}{2}\left(2{\bf {\dot R}}-
\frac{\hbar{\bf k}}{M}\right)
\times{\bf B}({\bf R)}\right]\right|^2
\label{20}
\eeq
where only transverse (i.e. solenoidal or divergence-free)  electromagnetic fields are involved and
we have written ${\bf E}$ instead of $-{\bn {\Pi}}/\epsilon_0$, anticipating free space quantisation [25].
The R\"ontgen term contains the average velocity of the atom before and
after the transition due to the symmetrised term in Eq.(\ref {16}) but we may
ignore the momentum of the photon since this is small compared to the initial
momentum of the atom.

The free-space normalised electromagnetic fields
can be obtained straightforwardly, remembering that we should identify two orthogonal wave polarisations. We choose the $z$-direction,
i.e. along ${\bf {\dot R}}$, as the axis along which there will be
either a magnetic field (i.e. transverse electric or TE) or an electric field
(i.e. transverse magnetic or TM). We can then write for a given wavevector ${\bf k}=(k,\theta,\phi)$
the following normalised electric and magnetic field vector amplitude functions
\bea
{\bf E}_{\bf k}^{TM}({\bf r})&=&
\left(\frac{\hbar\omega}{2V\epsilon_0}\right)^{1/2}
\left[\cos(\theta)\cos(\phi){\bf {\hat x}}+\cos(\theta)\sin(\phi){\bf {\hat y}}-
\sin(\theta){\bf {\hat z}}\right]e^{i({\bf k}.{\bf r}-\omega t)}
\nonumber \\
{\bf B}_{\bf k}^{TM}({\bf r})&=&
\left(\frac{\hbar\omega}{2V\epsilon_0 c^2}\right)^{1/2}
\left[\sin(\phi){\bf {\hat x}}-\cos(\phi){\bf {\hat y}}\right]
e^{i({\bf k}.{\bf r}-\omega t)}
\nonumber \\
{\bf E}_{\bf k}^{TE}({\bf r})&=&
\left(\frac{\hbar\omega}{2V\epsilon_0}\right)^{1/2}
\left[\sin(\phi){\bf {\hat x}}-\cos(\phi){\bf {\hat y}}\right]
e^{i({\bf k}.{\bf r}-\omega t)}
\nonumber \\
{\bf B}_{\bf k}^{TE}({\bf r})&=&
\left(\frac{\hbar\omega}{2V\epsilon_0 c^2}\right)^{1/2}
\left[\cos(\theta)\cos(\phi){\bf {\hat x}}+\cos(\theta)\sin(\phi){\bf {\hat y}}-
\sin(\theta){\bf {\hat z}}\right]e^{i({\bf k}.{\bf r}-\omega t)}
\nonumber \\
\label{21}
\eea
where $V$ is a normalisation volume.

Turning finally to the Fermi golden rule formula in the unprimed (laboratory) frame S, we find that
 the spontaneous emission rate can be written as
\beq
\Gamma=\frac{2\pi}{\hbar^2}\sum_{{\bf k},\lambda} |ME|^2
\delta(\omega-\omega'_0/\gamma-{\dot R} k \cos(\theta))=
\frac{V}{4\pi^2\hbar^2}\sum_\lambda\int_0^\pi d\theta \int_0^{2\pi} d\phi
\frac{{\omega'_0}^3|ME|^2}{\gamma^3c^3(1-{\dot R}\cos(\theta)/c)^4}
\label{22}
\eeq
where the sum over $\lambda$ denotes the summation over the two wave polarisations,
TM and TE, and the right hand side emerges after performing the integration
over ${\bf k}$ with the help of the delta function.
Substituting for the matrix element squared from Eq.(\ref{20}) and the electric and magnetic fields from
Eq.(\ref{21}), we find that
the spontaneous emission rate for a dipole parallel and perpendicular to the
velocity in the laboratory frame $S$ are given, respectively, by
\beq
\Gamma_\parallel=\Gamma'_0\int_{-1}^{1} dx \frac{3(1-x^2)}
{4\gamma^5(1-{\dot R}x/c)^4}
=\frac{\Gamma'_0}{\gamma}
\label{23}
\eeq
\beq
\Gamma_\perp=\Gamma'_0\int_{-1}^{1} dx \frac{3[(1+{\dot R}^2/c^2)(1+x^2)
-4{\dot R}x/c]}{8\gamma^3(1-{\dot R}x/c)^4}
=\frac{\Gamma'_0}{\gamma}
\label{24}
\eeq
where
$\Gamma'_0=d'^2 {\omega'_0}^3/(3\pi\epsilon_0 \hbar c^3)$ is the free-
space rate of spontaneous emission of the atom in the atomic rest-frame $S'$.
Note that only the TM mode is involved in the evaluation of the parallel dipole rate and that both
polarisations, TM and TE, are needed to obtain the perpendicular dipole
rate.
It is seen that there is no angular dependence,  i.e. the rate of spontaneous
emission is isotropic, and it does indeed vary like a relativistic clock.

\section*{Acknowledgments}

The authors are grateful to Professor G. Barton for useful
discussions and for providing a copy of Ref.[19]. C.R.B would like to thank
the EPSRC for financial support and L.G.B. would like to thank the University
of York for financail support. This work has been carried out under the EPSRC
Grant No. GR/M16313.

\section*{References}

\begin{enumerate}
\item M. Wilkens, Phys. Rev. Lett. {\bf 72}, 5 (1994)
\item H. Wei, R. Han and X. Wei, Phys.  Rev.  Lett.  {\bf 75}, 2071 (1995)
\item C. R. Hagen, Phys. Rev.  Lett.  {\bf 77}, 1656 (1996); H. Wei, R. Han and X. Wei, ibid 1657
\item G. Spavieri,  Phys.  Rev.  Lett.  {\bf 81}, 1533 (1998)
\item M. Wilkens, Phys. Rev. Lett. {\bf 81}, 1534 (1998)
\item U. Leonhardt and M. Wilkens, Europhys. lett. {\bf 42}, 365 (1998)
\item U. Leonhardt and P. Piwnicki,  Phys.  Rev. Lett. {\bf 82}, 2426 (1999)
\item C R Bennett, L G Boussiakou and M Babiker,  Phys. Rev. A, in press
\item E. Arimondo and H.-A. Bachor, eds. Quant.  Sem. Optics {\bf 8} (special issue), 495-753 (1996)
\item K. G. H. Baldwin,  Aust. J. Phys. {\bf 49}, 855-897 (1996)
\item C. S. Adams and E. Riis, Prog.  Quant.  Electr. {\bf 21}, 1 (1997)
\item M.  Babiker, E. A. Power and T. Thirunamachandran, Proc.  R.  Soc. Lond. A {\bf 332}, 187 (1973); {\bf 338}, 235 (1974)
\item E. A. Power and T. Thirunamachandran, Proc. R. Soc. Lond. A {\bf 372}, 265 (1980)
\item C. Baxter, M. Babiker and R. Loudon, Phys. Rev. A {\bf 47}, 1278 (1993)
\item V. E. Lembessis, M.  Babiker, C. Baxter and R. Loudon, Phys.  Rev. A {\bf 48}, 1594 (1993)
\item W. C. R\"ontgen, Ann. Phys. Chem. {\bf 35}, 264 (1888).
\item M. Wilkens Phys. Rev. A {\bf 47}, 671 (1993)
\item M. Wilkens, Phys. Rev. A {\bf 49}, 570 (1994)
\item G. Barton and A. Calogeracos in "The Cashimer Effect 50 Years Later",
Ed. M. Bordag (World Scientific: Singapore 1999)
\item M. Babiker and R. Loudon, Proc. R. Soc. Lond. A {\bf 385}, 439 (1983)
\item L. D. Landau and E. M. Lifshitz, "The classical theory of fields"
(Pergamon Press: Oxford 1969)
\item G. E. Vekstein Eur. J. Phys. {\bf 18}, 113 (1997)
\item Y. Aharonov and A. Casher, Phys. Rev. Lett. {\bf 53} 319 (1984)
\item A. P. French, "Special Relativity" (Chapman and Hall: UK 1990)
\item C. Cohen-Tannoudji, J. Dupont-Roc and G. Grynberg, "Photons and Atoms,
Introduction to Quantum Electrodynamics" (Wiley: New York 1989)
\end{enumerate}

\end{document}